 \documentstyle[epsfig,twocolumn,prb,aps]{revtex}

\begin{document}
\draft

 \twocolumn[\hsize\textwidth\columnwidth\hsize\csname @twocolumnfalse\endcsname

\title{
Topological effects at short antiferromagnetic Heisenberg chains
}
\author{
Jizhong Lou${}^{1}$, 
Shaojin Qin${}^{1}$, 
Tai-Kai Ng${}^{2}$, 
Zhaobin Su${}^{1}$
}
\address{
${}^{1}$Institute of Theoretical Physics, P O Box 2735, 
	Beijing 100080, P R China\\
${}^{2}$Department of Physics, 
	Hong Kong University of Science and Technology, Hong Kong
}
\date{ \today }
\maketitle

\begin{abstract}
The manifestations of topological effects in finite antiferromagnetic 
Heisenberg chains is examined by density matrix renormalization group 
technique in this paper. We find that difference between integer and 
half-integer spin chains shows up in ground state energy per site when
length of spin chain is longer than $\sim\xi$, where 
$\xi\sim\exp(\pi S)$ is a spin-spin correlation length, for spin 
magnitude S up to $5/2$.  For open chains with spin magnitudes $S=5/2$ 
to $S=5$, we verify that end states with fractional spin quantum 
numbers $S'$ exist and are visible even when the chain length is much 
smaller than the correlation length $\xi$.  The end states manifest 
themselves in the structure of the low energy excitation spectrum.  
\end{abstract}


]

\narrowtext

The importance of topological term in antiferromagnetic spin chains was
first pointed out by Haldane by mapping the Heisenberg model onto 
nonlinear $\sigma$-model (NL$\sigma$M)\cite{hald,aff0}. As a direct 
manifestation of the topological effect, Haldane pointed out that 
(infinite) integer spin chains have a gap in their excitation spectrum, 
whereas half-integer spin chains are gapless. The gap $\Delta$ in 
integer spin chains scale as $\Delta \sim Je^{-\pi{S}}$, where $J$ is 
the exchange coupling, and $S$ the spin magnitude. Topological effect 
is expected to manifest itself also at properties not directly related 
to excitation spectrum, for example, oscillation in ground state 
energies as $S$ changes from half-integer to integer\cite{zha}.  More 
recently Ng\cite{ng} pointed out that end states\cite{aff1} appear as 
lowest energy excitations in long open spin chains due to the same
topological term, and the structure of end states spectrum is determined 
by the spin magnitude $S$ of the spin chain\cite{ng}. In this paper we 
shall study numerically the manifestation 
of topological effect in finite spin chains with short lengths. 

Naively, we expect that difference between integer and half-integer 
spin chains becomes unobservable when $L\leq\xi\sim{e}^{\pi S}$, when 
the Haldane gap energy is comparable with the lowest spinwave 
excitation energy of the finite spin chain and becomes unobservable. If 
this is the case, then we expect that the topological effect will be 
hard to observe for spin chains with large value of spin magnitude $S$, 
where the correlation lengths are very long.  However, the derivation 
of the NL$\sigma$M assumes only that the length scale is much larger 
than lattice spacing and the topological term exists as long $L>>1$, 
independent of the correlation length $\xi$. Therefore, we expect that 
topological effect may manifest itself in some properties of spin chain 
even when chain length $L$ is less than $\xi$, as long as $L>>1$. In 
this paper, we shall study two properties of finite spin chains: the 
ground state energy and the end states. We find that oscillation in 
ground state energies between integer and half-integer spin chains is 
indeed observed numerically when chain length $L\geq\xi$. However the 
topological end states are much more robust and appear in finite spin 
chains even when $L<<\xi$. 

We use the DMRG method\cite{white} to calculate the ground state energy
per site for antiferromagnetic Heisenberg open chains with Hamiltonian
\begin{equation}
\label{sham}
H=\sum_{j=1}^{L-1}{\bf S}_{j}\cdot{\bf S}_{j+1},
\end{equation}
where the chain length is $L$ and ${\bf S}$ is spin operator. First we 
consider the large $L$ limit. In this limit the ground energy $e_0(S)$ 
can be obtained accurately as half of the difference of the energies for
length $L$ and $L+2$ chain in DMRG method\cite{huse}.  We keep $m=500$ 
states in DMRG, and the biggest truncation error is $10^{-6}$. 
We obtain the following converging results for thermodynamic limit 
from data of $L$ up to several thousands:
\vskip 0.1 truecm
\begin{tabbing}
111 \= 222222 \= 3333333333333  \= 222222 \= 33333333333333    \= \kill
\>  $S$ \>   $e_0(S)$    \>  $S$ \> $e_0(S)$     \\ \\
\>  0.5 \> -0.4431471    \>  1.0 \> -1.4014841   \\
\>  1.5 \> -2.8283304    \>  2.0 \> -4.76124365  \\
\>  2.5 \> -7.1922313    \>  3.0 \> -10.1237525  \\
\>  3.5 \> -13.5553061   \>  4.0 \> -17.486892    \\
\>  4.5 \> -21.918498    \>  5.0 \> -26.850118   
\end{tabbing}
The ground state energy for $S=1/2$ chain is exactly known, with 
$e_0(0.5)= 1/4-\ln(2)\sim-0.4431471$. The energy for $S=1$ chain has 
also been obtained to very high accuracy\cite{huse}, where 
$e_0(1)=-1.401484038971$.  In a paper demonstrating the $k=1$ $SU(2)$ 
WZW low energy behavior of $S=3/2$ chain\cite{hall}, the ground state 
for it has been accurately obtained, with $e_0(1.5)=-2.82833$.  The 
energy for $S=2$ chain was obtained in another paper demonstrating 
its massive relativistic low energy property\cite{xqw}, 
$e_0(2)=-4.761244$.  Our calculation is in agreement with these earlier 
studies\cite{lhq}.

\begin{figure}[ht]
 \epsfxsize=3.3 in\centerline{\epsffile{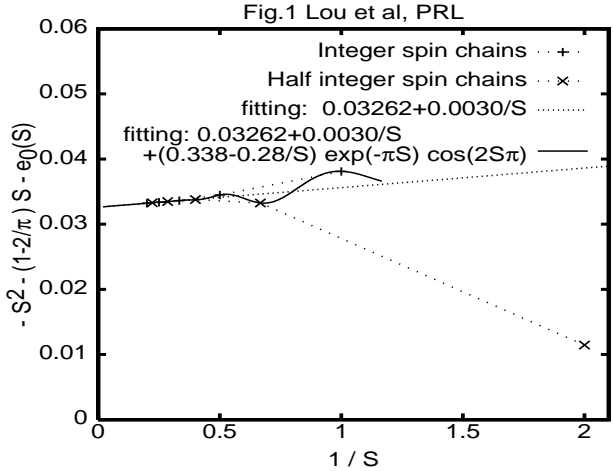}}
\vspace{0.5cm}
\caption[]{
Ground state energy for antiferromagnetic Heisenberg chains.
$-S^2-(1-2/\pi)S-e_0(S)$ vis $1/S$ is plotted and fit by polynomial of
$1/S$.  The ground state energies for integer spin chains and half
integer spin chains cannot be fitted by a single monotonic curve.  The 
difference decreases to zero very fast as $S$ increases. We note that
the $S=1/2$ point is not used in the fitting process
}
\label{fig1}
\end{figure}

The energies can be compared with a $1/S$ expansion:
\begin{equation}
e_0(S) = - S^2 + (2/\pi-1)S + a_0 + a_1/S + \ldots,
\end{equation}
where the first two terms were obtained from spin-wave theory\cite{and}.
In Fig. \ref{fig1} we plot $-S^2-(1-2/\pi)S-e_0(S)$ as a function of 
$1/S$. Our numerical result shows clearly the oscillatory nature of 
$e_0(S)$ between integer and half integer spin chains and $e_0(S)$ 
cannot be fitted by a single monotonic function of $1/S$. Assuming that 
the oscillatory behaviour is coming from topological terms in 
$NL\sigma{M}$ we expect that the oscillatory part of the ground state 
energy scales as $exp(-\pi S)$ and is nonanalytic in a 
$1/S$ expansion. This is indeed consistent with our numerical result as 
can be seen from the fitting in Fig. \ref{fig1}. 

\begin{figure}[ht]
 \epsfxsize=3.3 in\centerline{\epsffile{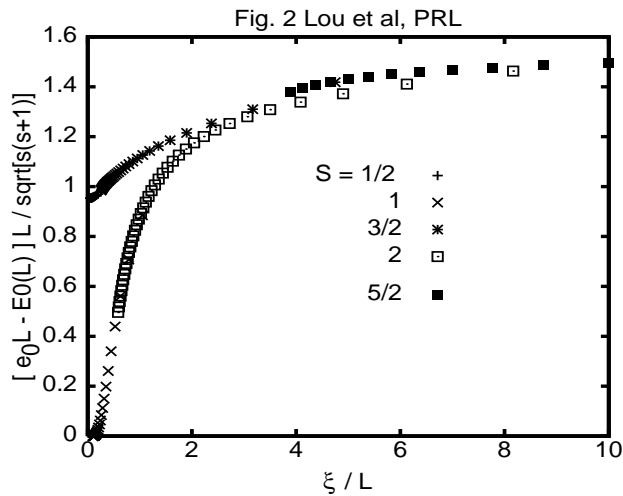}}
\vspace{0.5cm}
\caption[]{
Finite size scaling for ground state energy $E0(L)$ for 
antiferromagnetic Heisenberg chains. $(e_0-E0/L)/\sqrt{S(S+1)})$ 
versus inverse of the renormalized length $L/\xi$ is plotted for 
$S=1/2,1,3/2,2$, and $5/2$ with approximated values of $\xi=2,6.3,19,49$, 
and $140$, respectively.  Integer $S$ and half-integer $S$ chains behave 
differently in finite size scaling, but the difference vanishes for
small $L$ when $\xi/L\geq 1$,
}
\label{fig2}
\end{figure}

As is shown in Fig. \ref{fig1}, for $S\geq 3$, the energy differences 
between integer and half integer spin chains due to topological effect 
are very weak.  To examine the chain length dependence of ground state
energy we plot in Fig. \ref{fig2} the the length dependence of the 
finite length correction of the ground state energy 
$L [e_0 L-E0(L)]/\sqrt{S(S+1)}$ versus the inverse of renormalized 
length $(L/\xi)^{-1}$ for various value of spin magnitude $S$ up to $S=5/2$. 
The correlation lengths are estimated independently through various
fittings and the scaling behaviour is insensitive to fluctuations in
values of $\xi$ within $\sim10$ percent. $E0(L)$ is computed using DMRG method 
for chains with periodic boundary condition (PBC). It is clear from the figure 
that the ground state energies of integer and half-integer spin chains 
behave differently in finite size scaling, and the difference vanishes 
as $L\leq\xi$, in agreement with naive expectation based on Haldane gap 
argument.

It was pointed out in Ref. [\cite{ng}] that topological effect 
also leads to appearance of end states in open spin chains. For integer 
spin chains with length $L>>\xi$ end spins with magnitude $S/2$ appear 
at each end of spin chain and are coupled to each other with effective 
Hamiltonian
\begin{equation}
\label{eff}
H_{eff}=J(L) {\bf S'}_{1}\cdot{\bf S'}_{L},
\end{equation}
where ${\bf S'}_{1}$ and ${\bf S'}_{L}$ are the two end spins on the 
ends of the open chain.  $J(L)\sim{J}e^{-L/\xi}$ is positive for even 
$L$ and negative for odd $L$. For half-integer spin chains end spins
with magnitude $(S-1/2)/2$ appear and are coupled to each other with
same effective Hamiltonian (3) except that $J(L)\sim{J\over{L\ln(L)}}$ for 
large $L$. Notice that bulk spinwave excitations have energies scale as 
$\sim{1/L}$ for large $L$ and the lowest spin excitations of both 
integer and half-integer open spin chains are end-spin excitations when 
$L$ is large enough. Note that end states for $S=1$\cite{huse,ken} 
and $S=3/2,2$\cite{qin} open chains have been observed numerically in
long chains $L>>\xi$ with properties in agreement with end state 
theory\cite{ng}. The question of interests here is whether these end 
states remain robust as spin value $S$ increases, with length of spin 
chains reduce to $L<\xi$. We expect that the end states may stay robust 
because in general energy level crossings have to occur if end 
states are moved out of the low energy excitation spectrum.  We shall 
show in the following that end states are observed as lowest energy 
excitations in open spin chains for large spin magnitude $S$ up to 5 
with chain lengths much smaller than $\xi$. 

%
%

%
%
  
We consider spin chains with spin magnitudes $5/2 \leq S\leq 5$ 
and with even number of sites with chain lengths from $L=4$ to $30$. 
Notice that by restricting ourselves to chains with large value of $S$ 
and short lengths $\leq 30$ we are always in the limit $L\leq\xi$ in 
our numerical study. By keeping $m=450$ states in DMRG.  We calculate 
the lowest energy states in sectors with $S^{tot}_z=0,1,\ldots,S+1$.  
The biggest truncation error is $10^{-4}$ for $S=5$. According to end 
state theory, the lowest $2S'+1$ states in the excitation spectrum 
should have $S^{tot}=0$, $1$, \ldots, $2S'$ in increasing order of 
energy, where $S'=S/2$ for integer spin chains and $S'=(S-1/2)/2$ for 
half-integer chains, the bulk spinwave spectrum appears only above 
these states.
  
\begin{figure}[ht]
 \epsfxsize=3.3 in\centerline{\epsffile{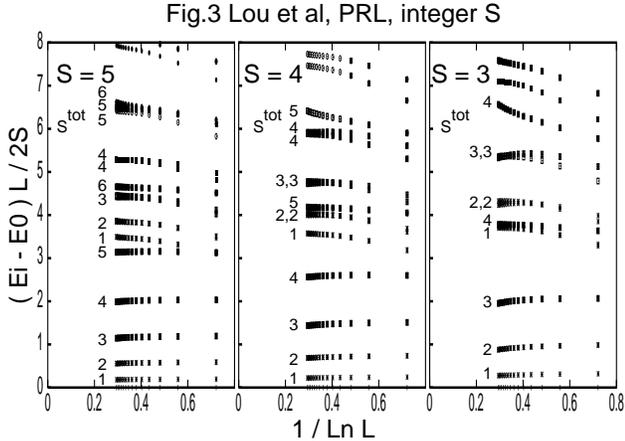}}
\vspace{0.5cm}
\caption[]{
Low energy excitations times chain length for integer spin $S=3$ to 
$5$ open chains.  The ground state is of total spin $S^{tot}=0$.  The 
$S+1$ lowest excited states from low energy to high energy are 
$S^{tot}=1,\ldots,S$, and $S^{tot}=1$ from bottom to top.
}
\end{figure}

\begin{figure}[ht]
 \epsfxsize=3.3 in\centerline{\epsffile{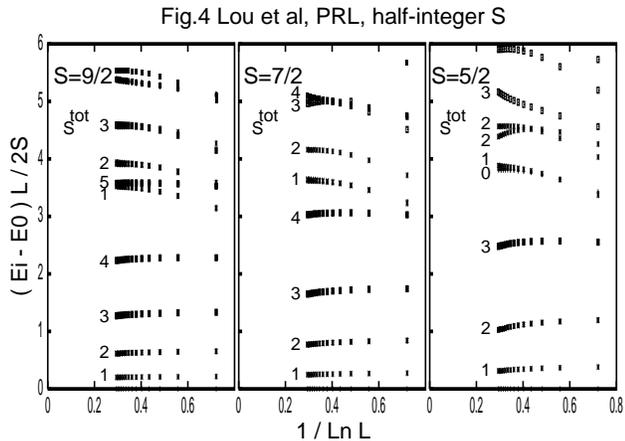}}
\vspace{0.5cm}
\caption[]{
Low energy excitations times chain length for half-integer spin 
$S=5/2$ to $S=9/2$ open chains.  The ground state is of total spin 
$S^{tot}=0$.  The $S-1/2$ lowest excited states from low energy to 
high energy are $S^{tot}=1,\ldots,S-1/2$.
}
\end{figure}

We show the low energy excitation spectrum for open integer spin chains 
in Fig. 3.  Starting from the ground state energy $E0$, we label the 
$i^{th}$ excited states' energy as $Ei$.  For integer spin chains, we
find indeed that the lowest $S+1$ energy levels are ordered in 
sequence with $S^{tot}=0,1,\ldots,S$ from ground state to the $S^{th}$ 
excited states. Above this series of states, spinwave excitations appear, 
starting with total spin $S^{tot}=1$. The corresponding low energy 
spectrum for open half integer spin chains is shown in Fig. 4, and
the same structure is observed- the lowest $S+1/2$ energy levels 
are ordered in sequence with $S^{tot}=0,1,\ldots,S-1/2$ from ground state 
to the $(S-1/2)^{th}$ excited states. Above these series of states, 
spinwave excitations appear. These are exactly the spectra predicted by 
the end state theory. 

To examine the nature of these low lying states further we investigate 
the validity of Eq. (\ref{eff}) in describing the energies of these 
states.  It is easy to show that Eq.\ (\ref{eff}) predicts that the 
number series $y(i)=(Ei - E0)/(E1 - E0)$ for the end states is given by 
the simple mathematical expression $y(i)=i(i+1)/2$, independent of 
chain lengths $L$ and spin magnitude $S$.
 
\begin{figure}[ht]
 \epsfxsize=3.3 in\centerline{\epsffile{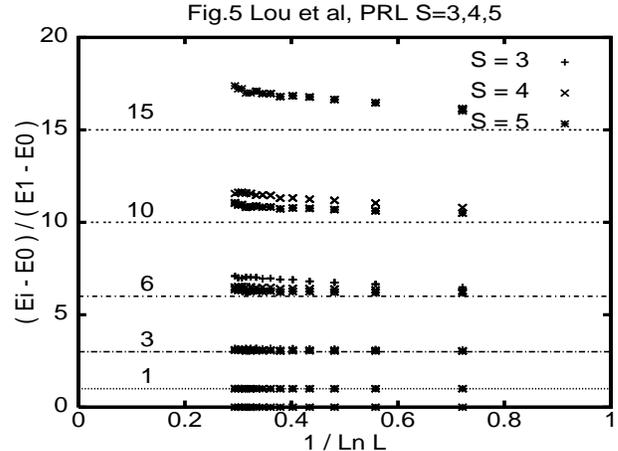}}
\vspace{0.5cm}
\caption[]{
Edge excitation energy for integer spin chains.
}
\
\end{figure}

\begin{figure}[ht]
 \epsfxsize=3.3 in\centerline{\epsffile{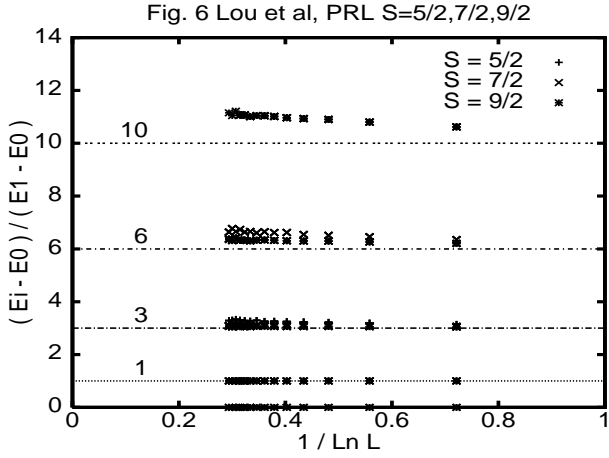}}
\vspace{0.5cm}
\caption[]{
Edge excitation energy for half integer spin chains.
}
\
\end{figure}

In Fig. 5 and Fig. 6, we plot respectively, $y(i)$ for the $2S'$ lowest 
energy levels given by DMRG versus $1/\ln L$ for integer spin chain and 
half integer spin chains. It is clear that Eq. (\ref{eff}) describes
quite well the qualitative behaviour of the lowest $2S'$ states of the
excitation spectra for all chains with different spin values and 
lengths under investigation, despite that we are already in the
$L<\xi$ limit. 

Summarizing, we study in this paper using DMRG method several
manifestions of topological effect in finite Heisenberg spin chains.
We confirm the oscillation of ground state energies between
integer and half-integer spin chains. The magnitude of the oscillatory
part of ground state energies goes down very rapidly as $S$ increases 
and is consistent with an $e^{-\pi S}$ behaviour. The oscillation becomes 
unobservable when chain length $L$ decreases to less 
than $\xi$, in agreement with argument based on Haldane gap. We have 
also study open spin chains where topological effect also manifest 
itself as end states.  Surprisingly, we find strong numerical evidence 
that end states stay robust as chain length $L$'s decrease and are 
observable even when $L<<\xi$. Our result confirms the theoretical 
expectation that the topological effect is a robust property of quantum 
spin chains and exists as long as chain length $L>>1$, independent of 
the appearance of Haldane gap in the excitation spectrum. 
 
This work is partially supported by Chinese Natural Science Foundation, 
HKUGC under RGC grant HKUST6143/97P. T.K. Ng acknowledge the 
hospitality of the Issac Newton Institute for Mathematical Sciences at 
which this paper is partly written.

\end{document}